\documentclass{appolb}
\usepackage{graphicx}
\usepackage{amsmath}
\usepackage{amssymb}

\begin{document}
\title{Investigating the onset of deconfinement with NA61/SHINE%
\thanks{Presented at the XXXII Cracow Epiphany Conference}%
}
\author{Oleksandra Panova\\for the NA61/SHINE Collaboration
\address{Institute of Nuclear Physics, Polish Academy of Sciences}
}
\maketitle
\begin{abstract}
NA61/SHINE is a multipurpose fixed-target experiment located at the CERN SPS. One of its main goals is to study the onset of deconfinement and the properties of strongly interacting matter. For this purpose, a unique two-dimensional scan in collision energy ($\sqrt{s_\mathrm{NN}} = 5.12 - 16.8/17.3$~GeV) and system size (from $p$+$p$ to Pb+Pb) was performed. Results on hadron spectra produced in nucleus-nucleus collisions, including the recent data on charged hadrons produced in central Xe+La collisions and baryons in central Ar+Sc collisions, are presented. The kinematic distributions and the measured multiplicities of identified hadrons are compared with NA49 Pb+Pb results and with available world data. The obtained results, particularly the $K^+/\pi^+$ ratio, are crucial for understanding the phenomena of the onset of deconfinement, which is one of the main aims of the strong interaction program of the NA61/SHINE Collaboration. Additionally, a comparison of proton rapidity spectra in nucleus-nucleus collisions from NA61/SHINE and NA49 is presented, providing a complete picture of the energy and system-size dependence of the mechanism of transport of baryon number at SPS energies.

\end{abstract}
  
\section{Introduction}
NA61/SHINE is a fixed-target experiment~\cite{NA61, NA61_1} located at the CERN SPS North Area. It is a multipurpose spectrometer measuring particle production in hadron-hadron, hadron-nucleus, and nucleus-nucleus collisions at beam momenta from 13$A$ to 400$A$ GeV/$c$ corresponding to center-of-mass energies per nucleon pair from 5.12 to 27.4~GeV. The goals of the experiment include the search for the critical point of strongly interacting matter and the study of the properties of the onset of deconfinement -- the beginning of the creation of quark-gluon plasma in nucleus-nucleus collisions with increasing collision energy. To reach these goals, the two-dimensional scan in collision energy and system size ($p$+$p$, $p$+Pb, Be+Be, Ar+Sc, Xe+La, Pb+Pb) was performed.

This program was mainly motivated by the observation of rapid changes of hadron production properties in central Pb+Pb collisions at about $30A$ GeV/$c$ ($\sqrt{s_\mathrm{NN}} = 7.62$~GeV) by the NA49 experiment~\cite{paper:PbPb1}: a sharp peak in the $K^+/\pi^+$ ratio (``horn''), the start of a plateau in the inverse slope parameter $T$ characterizing the kaon transverse momentum ($p_\mathrm{T}$) distribution (``step''), and a steepening of the increase of pion production per wounded nucleon with increasing collision energy (``kink''). These phenomena were earlier predicted as signatures of the onset of deconfinement~\cite{paper:smes}.  

The article is organized as follows. 
In Sec.~\ref{sec:methods}, the procedures of the particle identification are described. The experimental results obtained by NA61/SHINE are presented in Sec.~\ref{sec:res}, along with a comparison with available world data. Section \ref{sec:onset} discusses the collision energy and system size dependence of the obtained results. 

\section{Analysis methodology}
\label{sec:methods}
The main method of particle identification used by NA61/SHINE, the $\mathrm{d}E/\mathrm{d}x$ method, is based on the simultaneous measurement of the specific ionization energy loss of charged particles in the Time Projection Chambers (TPCs) together with their reconstructed momentum and charge. It allows us to identify all the most abundant charged hadrons: $\pi^-$, $\pi^+$, $K^-$, $K^+$, $p$, and $\bar p$. The $\mathrm{d}E/\mathrm{d}x$ method has limited acceptance -- it allows the identification of particles only with laboratory momentum $p_\mathrm{lab} \lesssim 1$~GeV/$c$ and $p_\mathrm{lab} \gtrsim 5$~GeV/$c$, which limits the acceptance at mid-rapidity. 
To improve the particle identification capabilities of the $\mathrm{d}E/\mathrm{d}x$ method, a supplementary measurement using the Time-of-Flight (ToF) detector is performed. It allows for  particle identification using the $tof - \mathrm{d}E/\mathrm{d}x$ method, which is based on the measurement of the $\mathrm{d}E/\mathrm{d}x$ from TPCs and the mass of the particle based on ToF and TPC data. Simultaneous analysis using the $\mathrm{d}E/\mathrm{d}x$ and $tof - \mathrm{d}E/\mathrm{d}x$ methods provides coverage of almost the full forward rapidity hemisphere for charged particles. 
More details about $\mathrm{d}E/\mathrm{d}x$ and $tof - \mathrm{d}E/\mathrm{d}x$ methods can be found in Refs.~\cite{paper:na61_p_dedx, paper:na61_be_dedx, paper:na61_arsc_dedx}.

Additionally, for the measurement of $\pi^-$ mesons, the $h^-$ method can be used~\cite{paper:na61_p_h_minus, paper:na61_be_h_minus, paper:na61_arsc_h_minus}.
This method is based on the fact that more than 90\% of primary negatively charged hadrons produced in heavy-ion interactions at the SPS energy range are $\pi^-$ mesons. Therefore, spectra of $\pi^-$ mesons may be reliably obtained by subtracting the small contribution of other particles from the spectra of all negatively charged hadrons ($h^-$) using Monte Carlo simulations tuned to experimental data.

Neutral particles leave no direct tracks in the detector; therefore, their identification relies on the reconstruction of their weak decay topologies. The method relies on identifying $\text{V}^0$ candidates originating from decays such as $\varLambda \rightarrow p + \pi^-$. More details concerning neutral particle identification in the NA61/SHINE experiment can be found in Ref.~\cite{paper:lambda_pp}. 
    
\section{Particle spectra}
\label{sec:res}

\begin{figure}[h]
        \begin{center}
        \includegraphics[trim=0 0.0cm 0 0.10cm,clip,width = 0.32\linewidth]{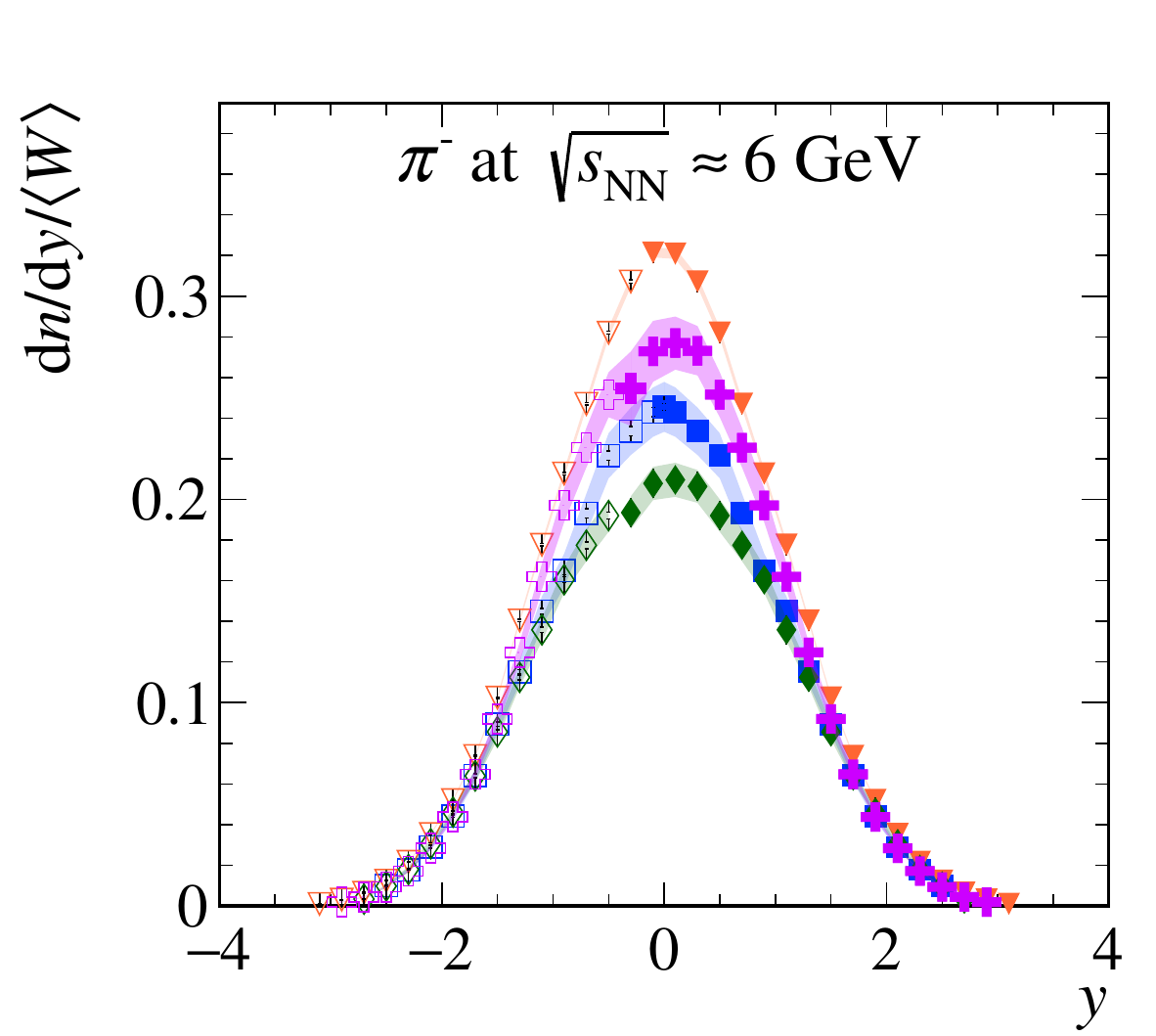}
        \includegraphics[trim=0 0.0cm 0 0.10cm,clip,width = 0.32\linewidth]{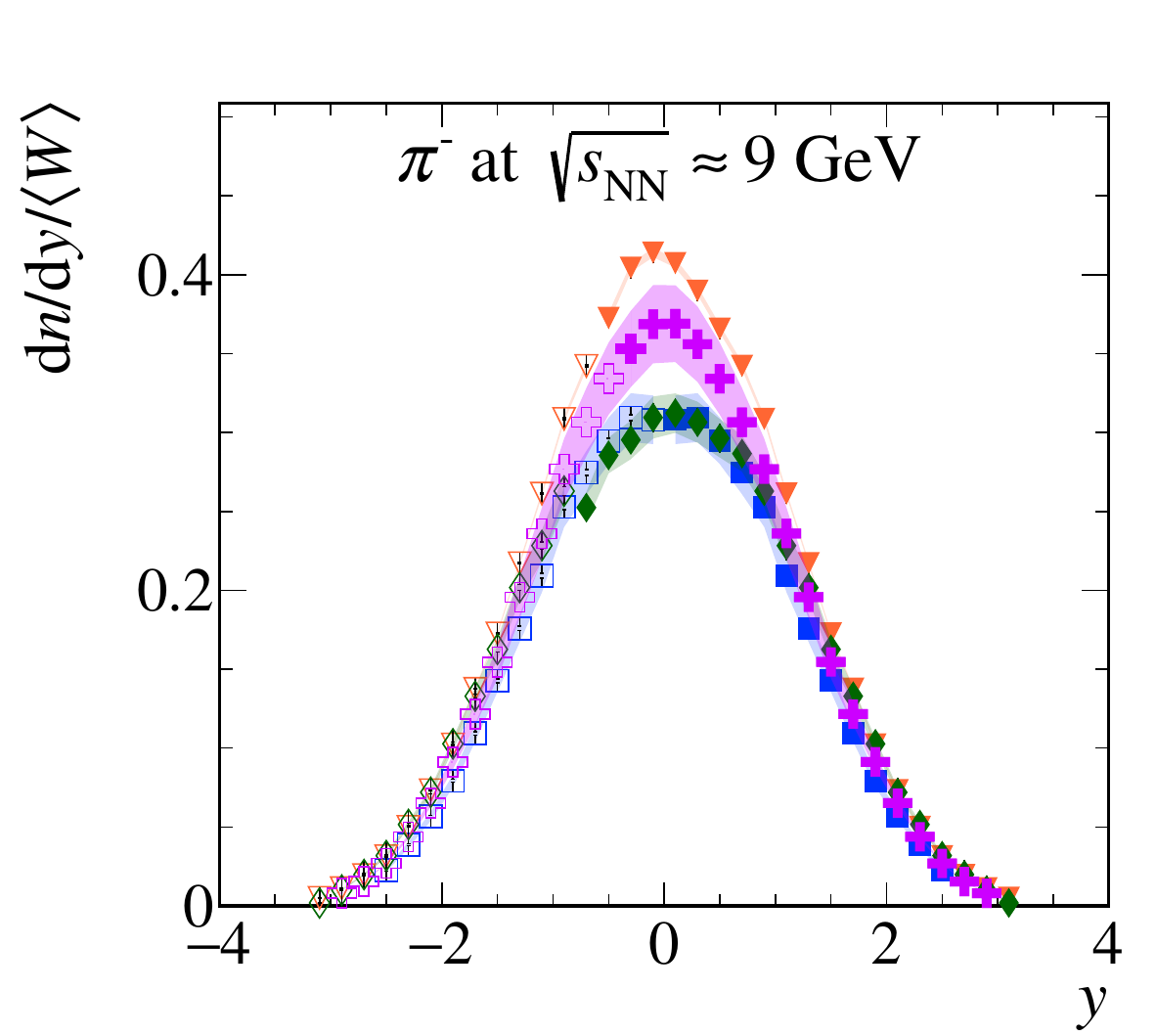}
        \includegraphics[trim=0 0.0cm 0 0.10cm,clip,width = 0.32\linewidth]{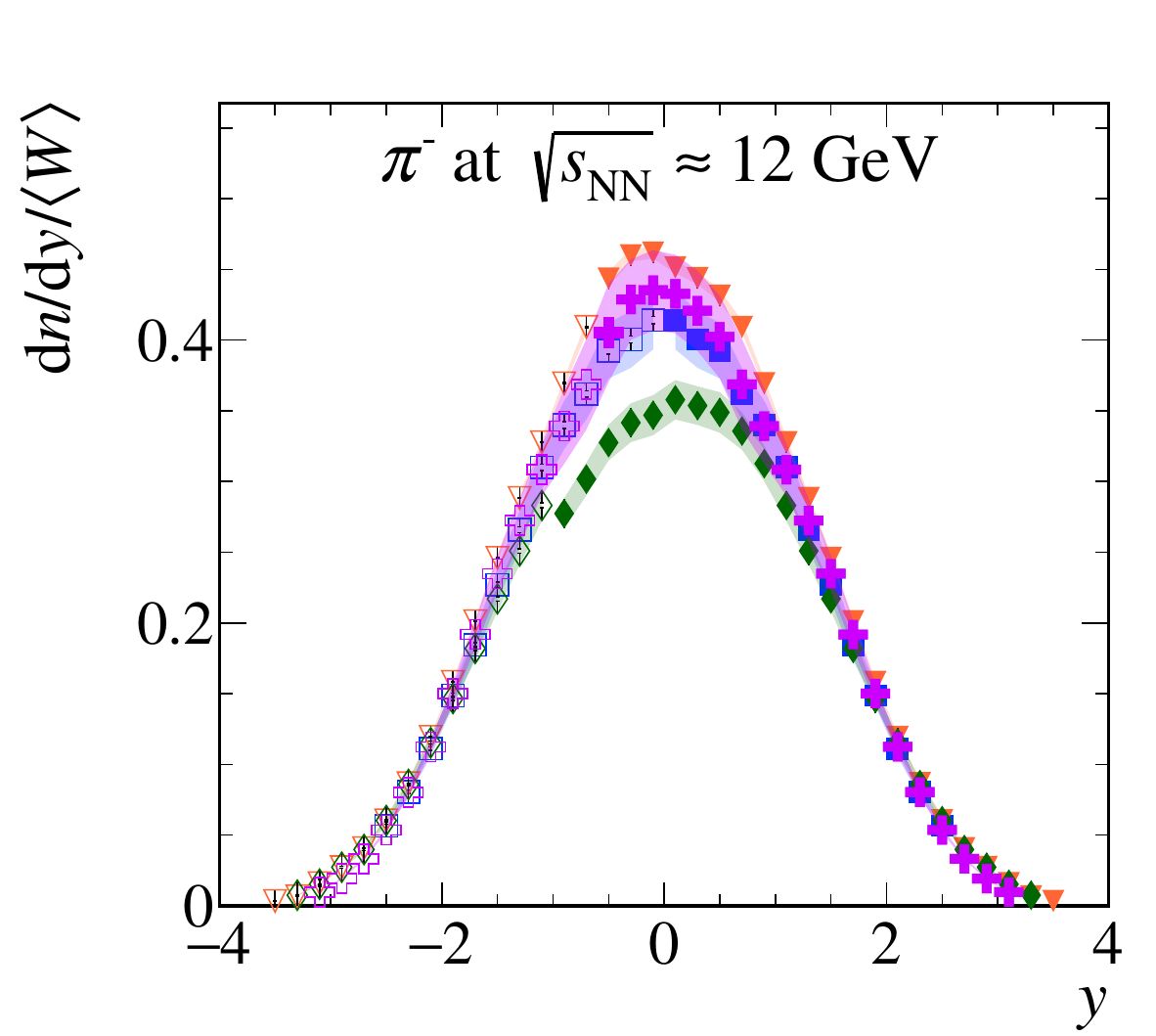}
        \vspace{-0.20cm}
        \includegraphics[width = 0.79155\linewidth]{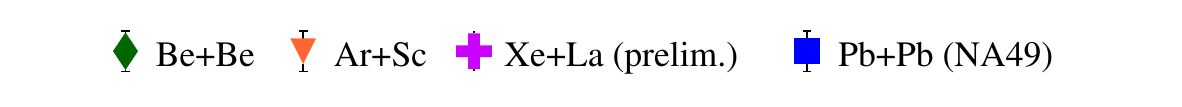}
        \caption{Comparison of rapidity distributions $\mathrm{d}n/\mathrm{d}y$ of $\pi^-$ mesons produced in central Be+Be, Ar+Sc, Xe+La, and Pb+Pb collisions at $\sqrt{s_\mathrm{NN}} \approx 6, 9,$ and $12$~GeV. Statistical uncertainties are shown with error bars, and systematic uncertainties are shown as shaded bands. Published data from Refs.~\cite{bebe1,paper:na61_arsc_h_minus,paper:PbPb3,paper:PbPb1}.}
        \label{figure:pi_minus}
\vspace{0.2cm}

        \includegraphics[trim=0 0.0cm 0 0.10cm,clip,width = 0.32\linewidth]{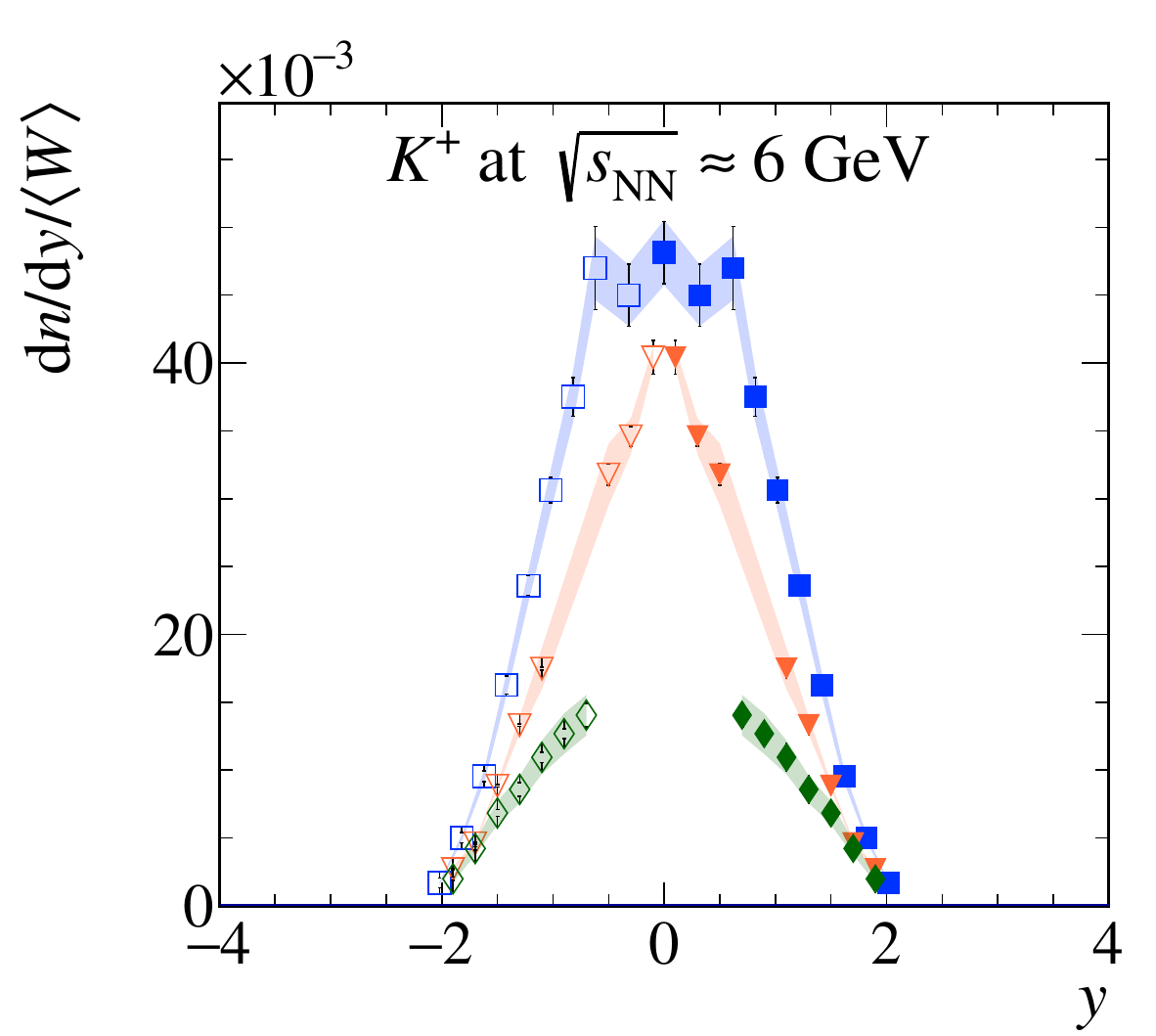} 
        \includegraphics[trim=0 0.0cm 0 0.10cm,clip,width = 0.32\linewidth]{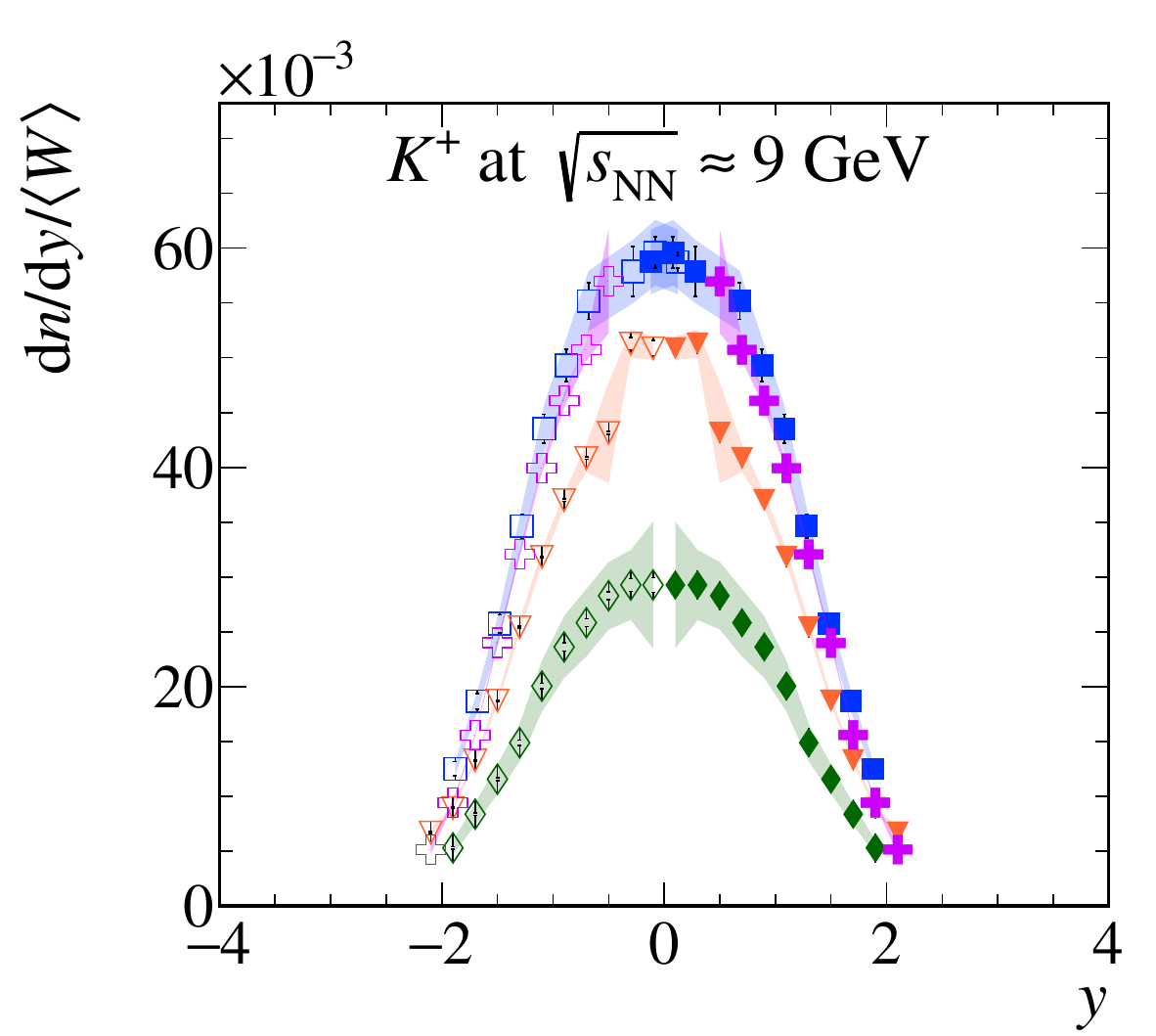}
        \includegraphics[trim=0 0.0cm 0 0.10cm,clip,width = 0.32\linewidth]{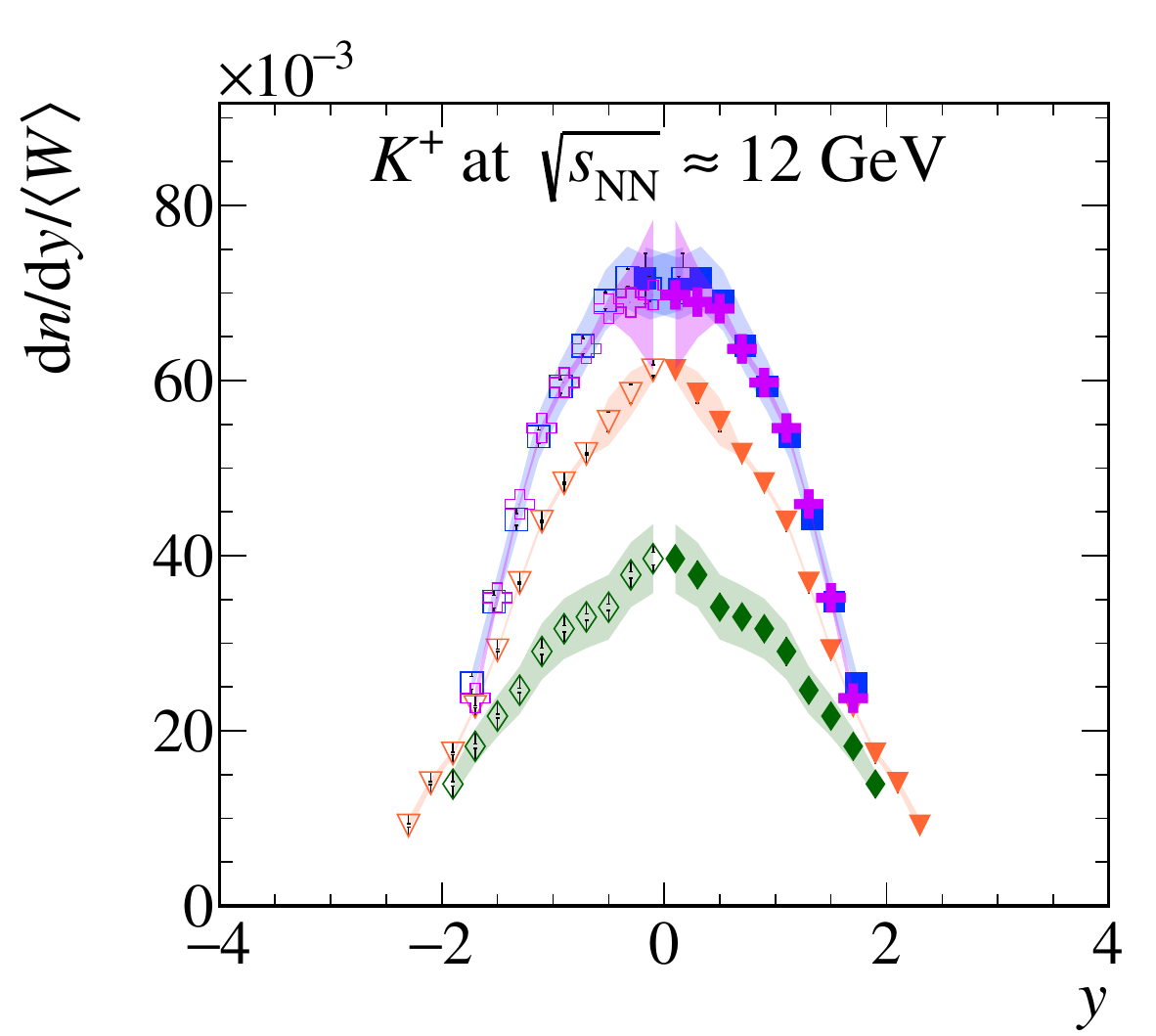} 
        \vspace{-0.15cm}
        \includegraphics[width = 0.79155\linewidth]{Xe_La_legend_systems_dndy.pdf}
        \caption{Comparison of rapidity distributions $\mathrm{d}n/\mathrm{d}y$ of $K^+$ mesons produced in central Be+Be, Ar+Sc, Xe+La, and Pb+Pb collisions at $\sqrt{s_\mathrm{NN}} \approx 6, 9,$ and $12$~GeV. Statistical uncertainties are shown with error bars, and systematic uncertainties are shown as shaded bands.  Published data from Refs.~\cite{bebe2,paper:na61_arsc_dedx,paper:PbPb3,paper:PbPb1}.}
        \label{figure:K}
        \end{center}
    \end{figure}

\begin{figure}[h]
        \begin{center}
        \includegraphics[trim=0 0.0cm 0 1.5cm,clip,width = 0.301\linewidth]{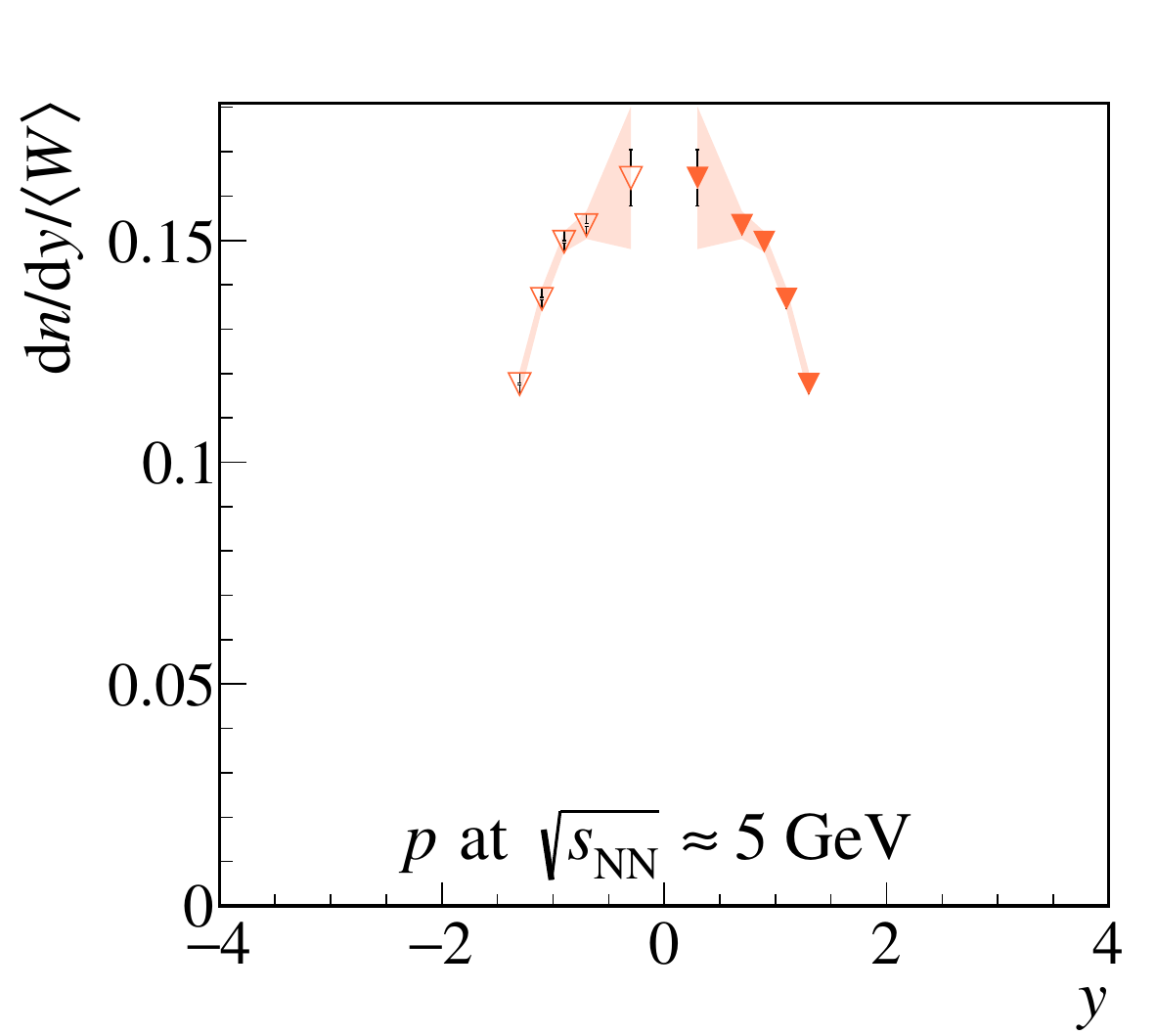} 
        \includegraphics[trim=0 0.0cm 0 1.5cm,clip,width = 0.301\linewidth]{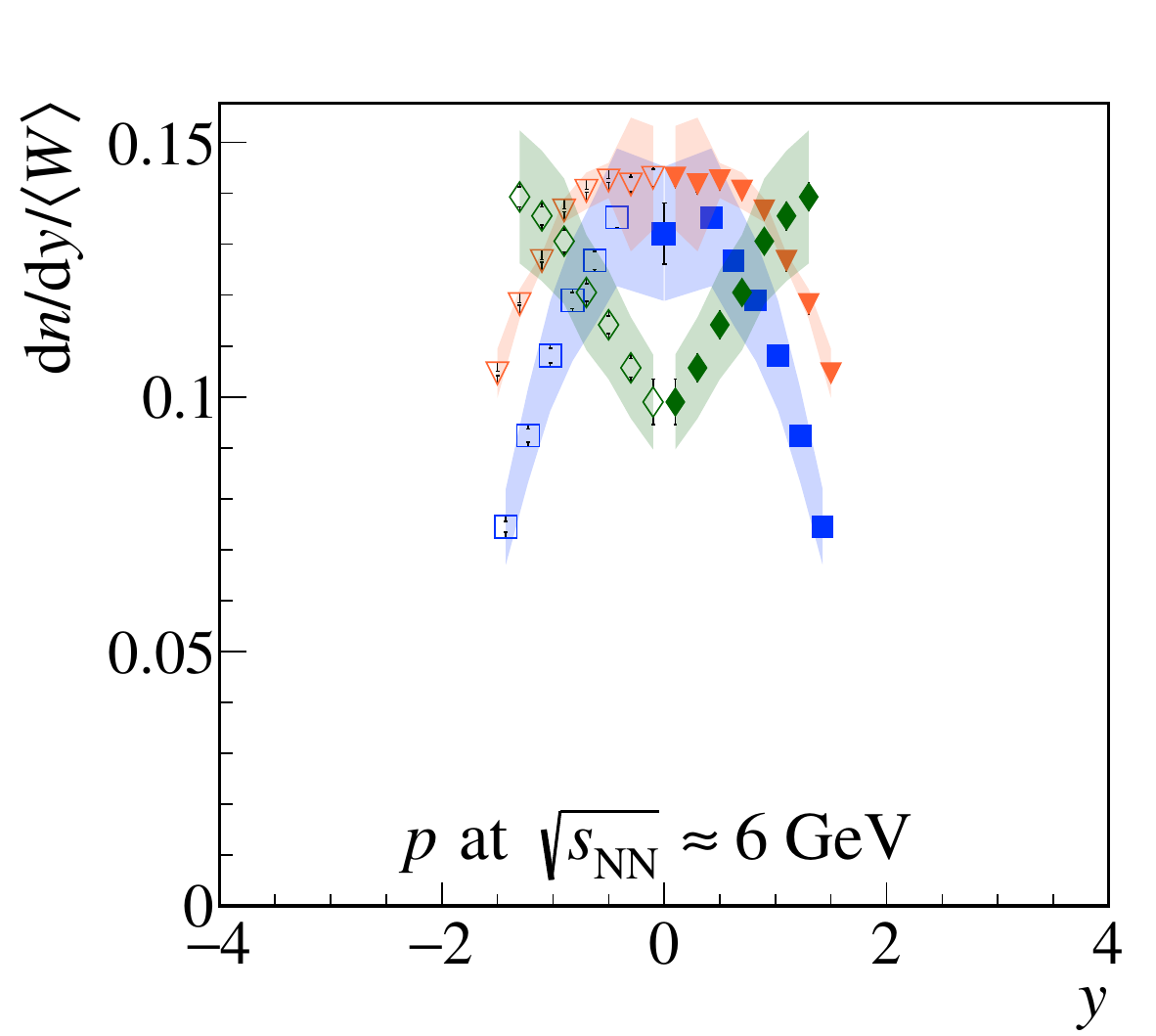} 
        \includegraphics[trim=0 0.0cm 0 1.5cm,clip,width = 0.301\linewidth]{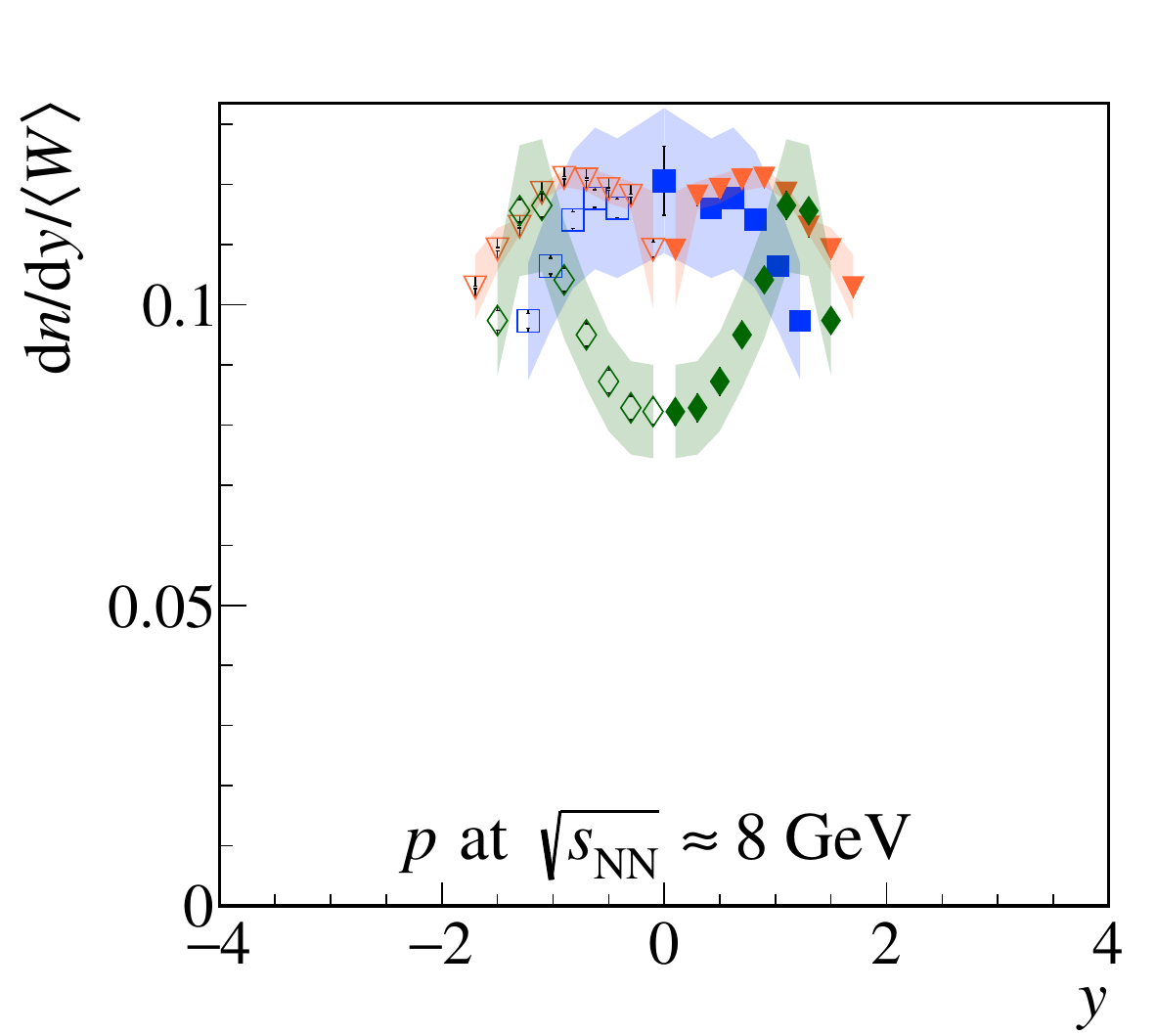} 
        \includegraphics[trim=0 0.0cm 0 1.5cm,clip,width = 0.301\linewidth]{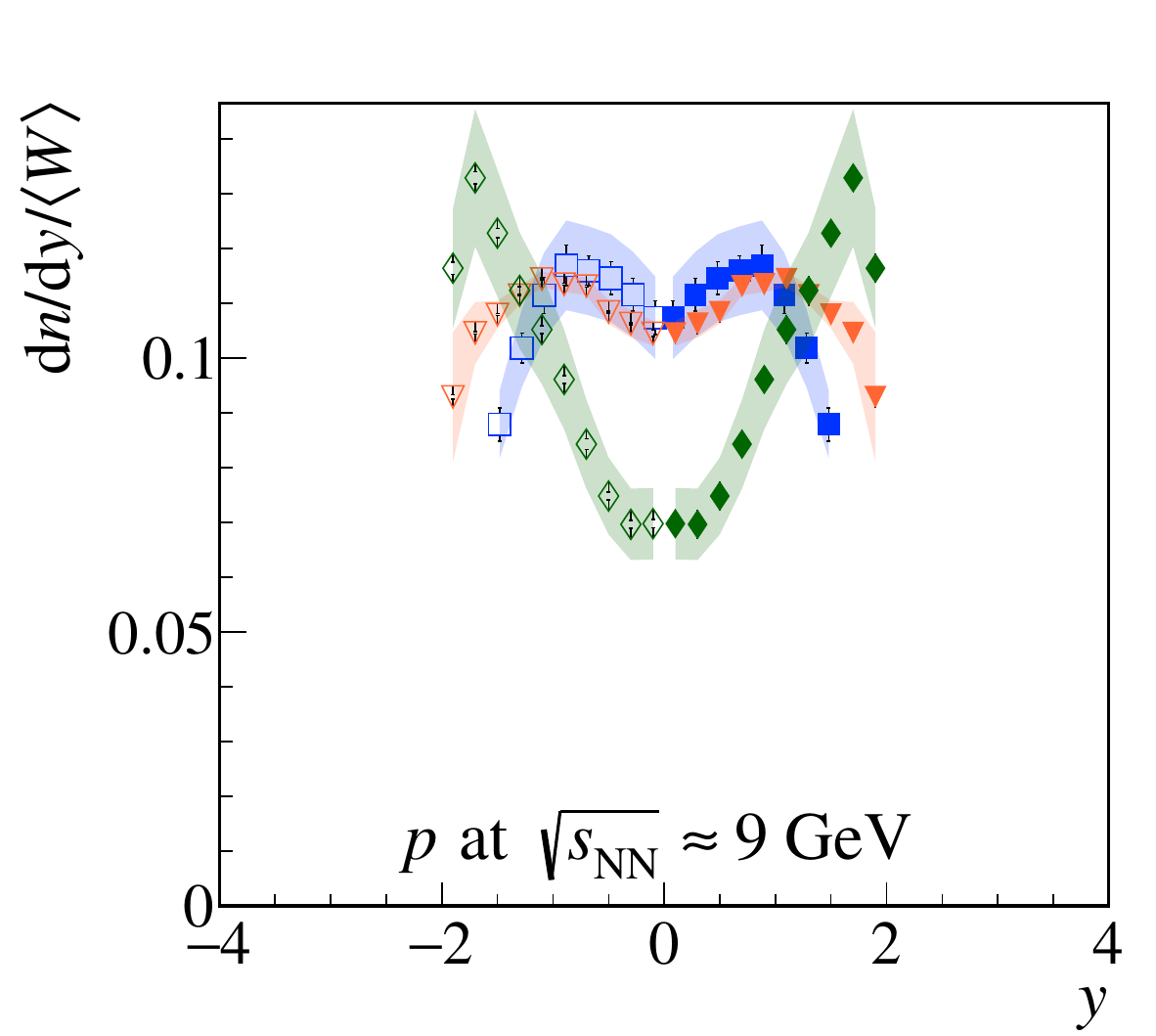} 
        \includegraphics[trim=0 0.0cm 0 1.5cm,clip,width = 0.301\linewidth]{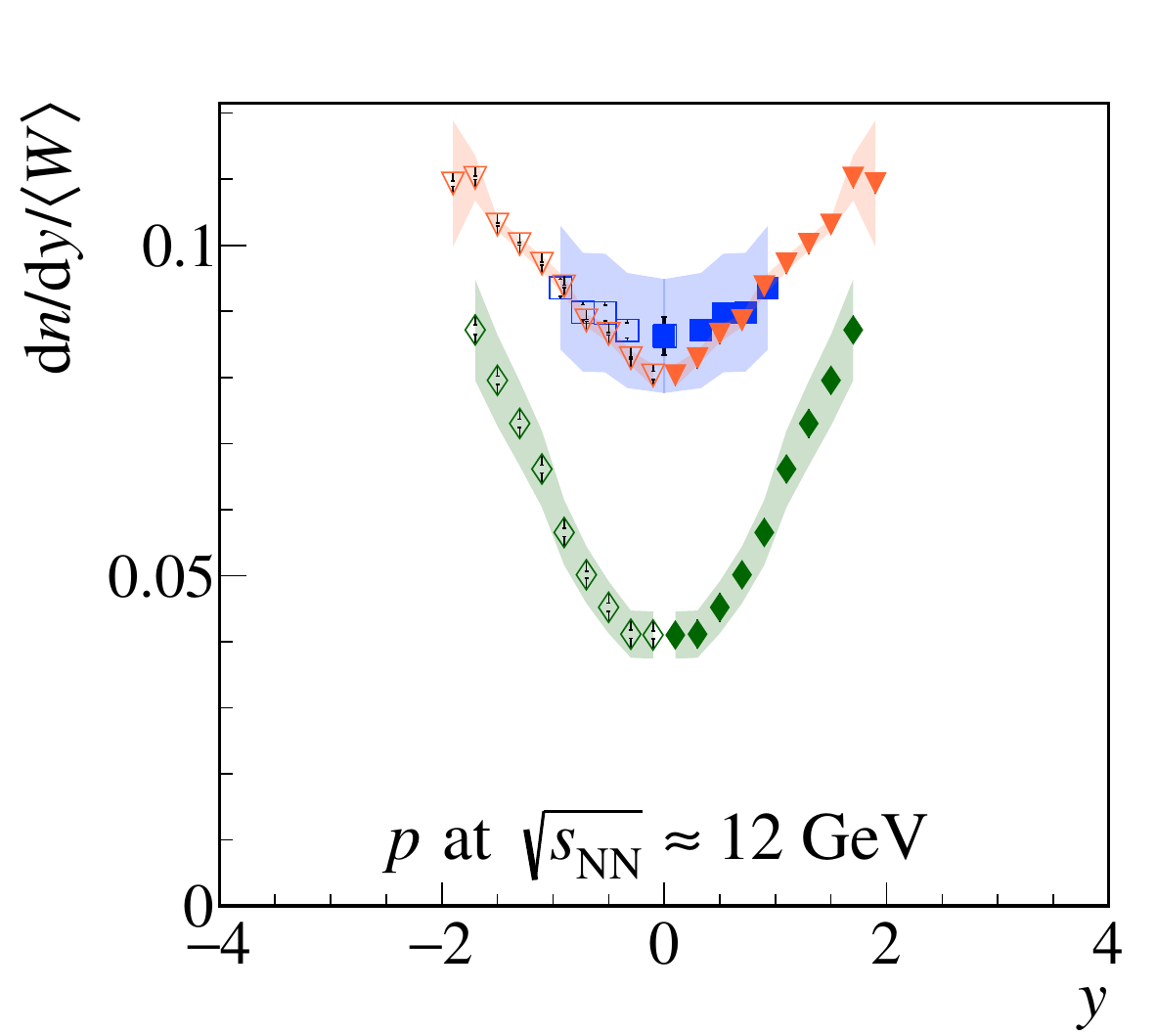} 
        \includegraphics[trim=0 0.0cm 0 1.5cm,clip,width = 0.301\linewidth]{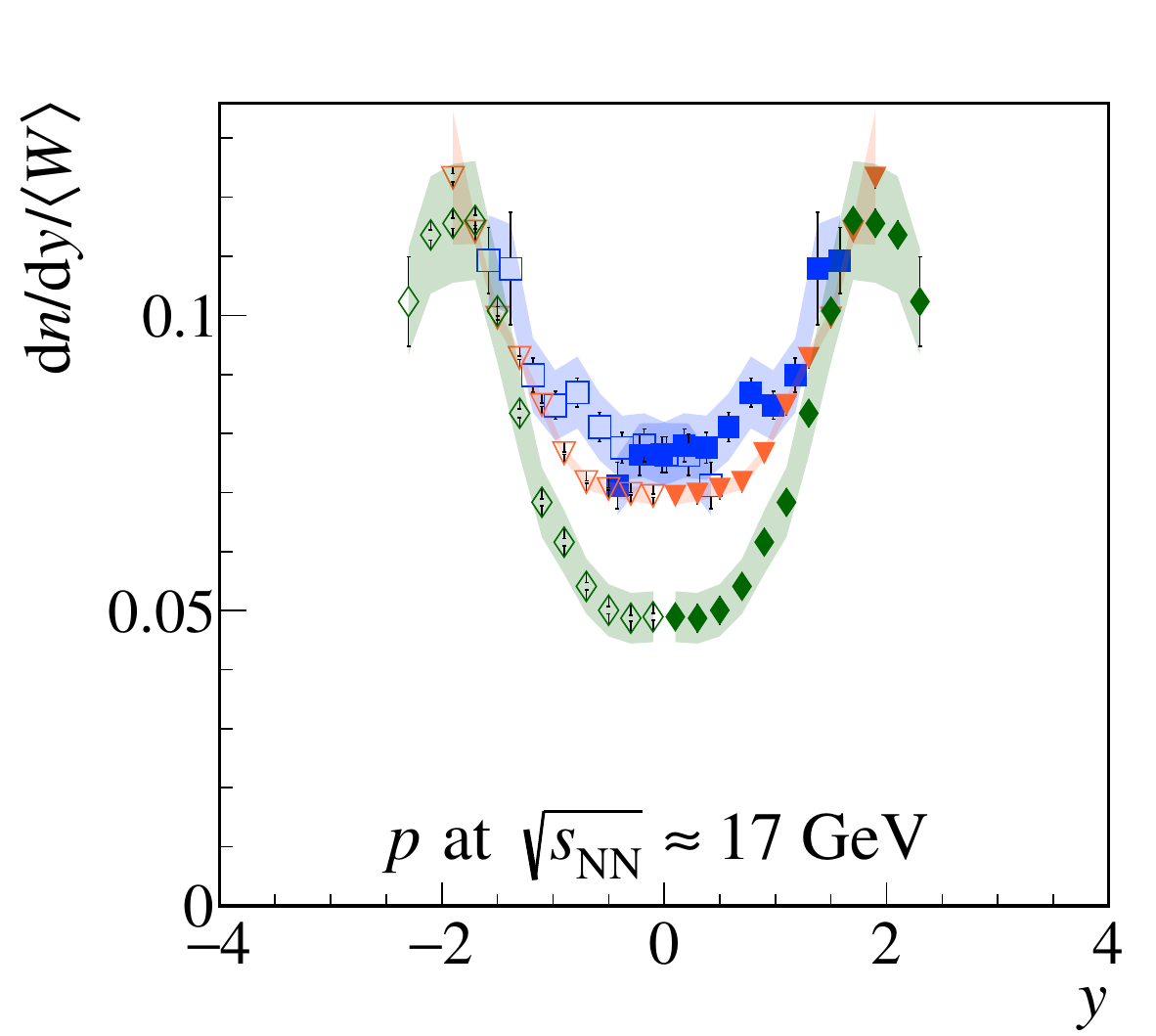} 
        \vspace{-0.35cm}
        \includegraphics[trim=0 0.0cm 2.7cm 0.0cm,width = 0.525155\linewidth]{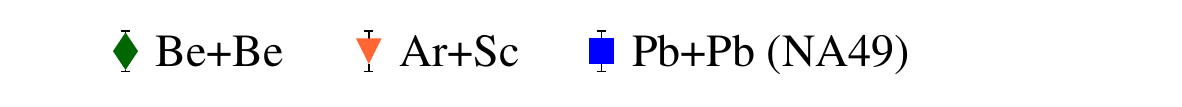}
        \caption{Comparison of rapidity distributions $\mathrm{d}n/\mathrm{d}y$ of protons produced in central Be+Be, Ar+Sc, and Pb+Pb collisions at $\sqrt{s_\mathrm{NN}} \approx 5, 6, 8, 9, 12,$ and $17$~GeV. Statistical uncertainties are shown with error bars, and systematic uncertainties are shown as shaded bands. Published data from Refs.~\cite{bebe2,paper:na61_arsc_dedx,na49_proton, na49_blume}.}
        \label{figure:p}
\vspace{0.2cm}
        \includegraphics[trim=0 0.0cm 0 1.5cm,clip,width = 0.301\linewidth]{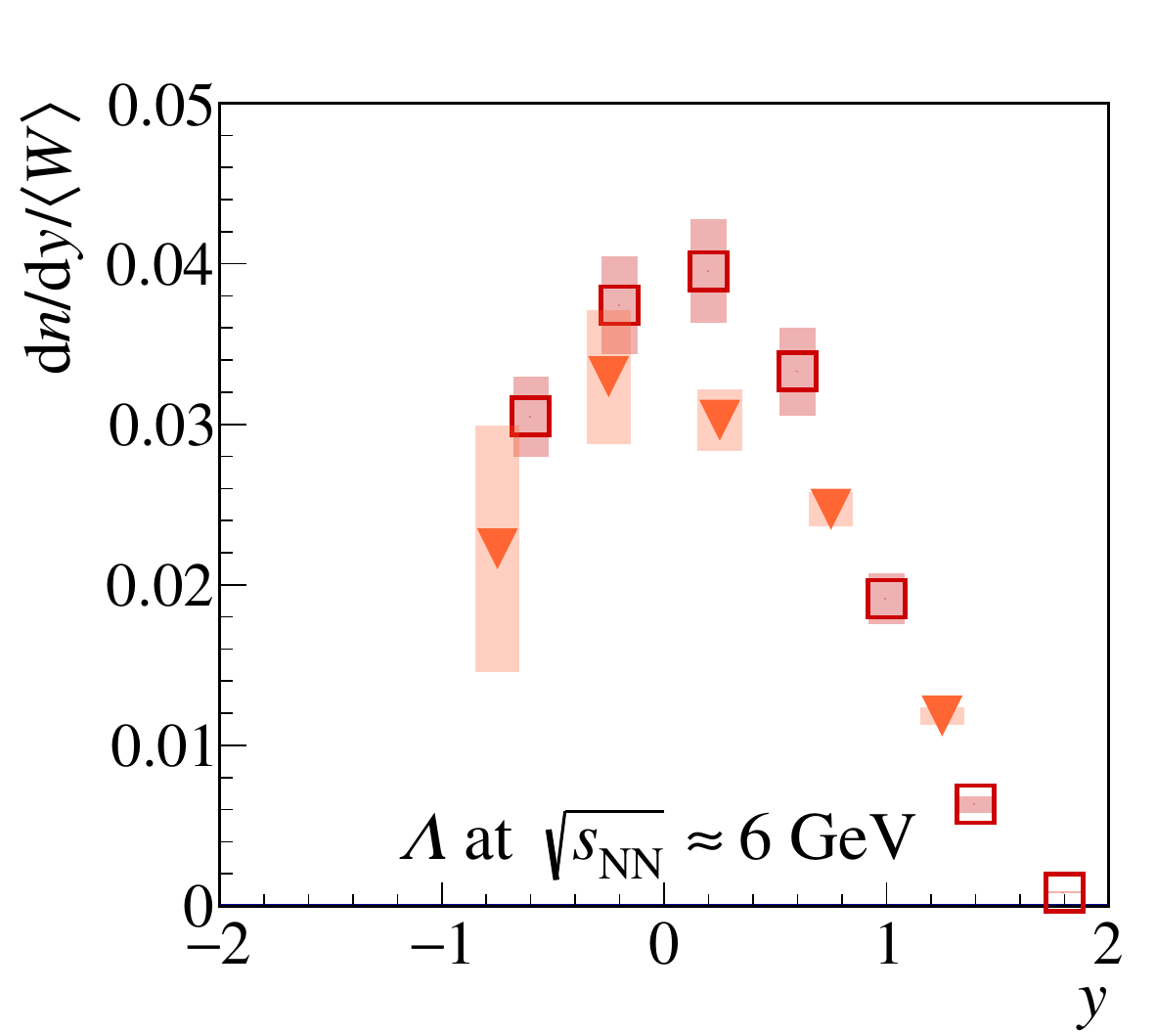} 
        \includegraphics[trim=0 0.0cm 0 1.5cm,clip,width = 0.301\linewidth]{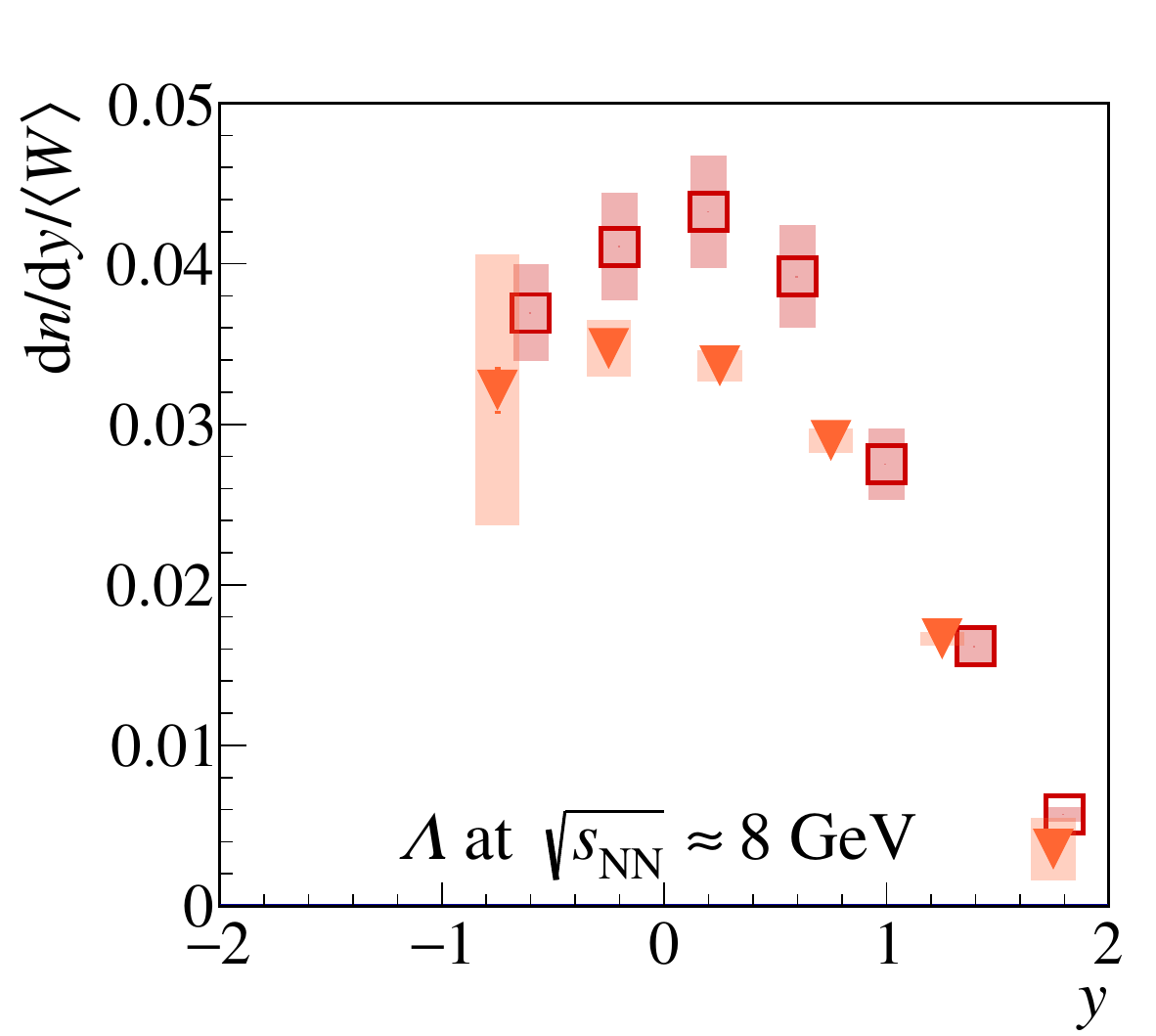} 
        \includegraphics[trim=0 0.0cm 0 1.5cm,clip,width = 0.301\linewidth]{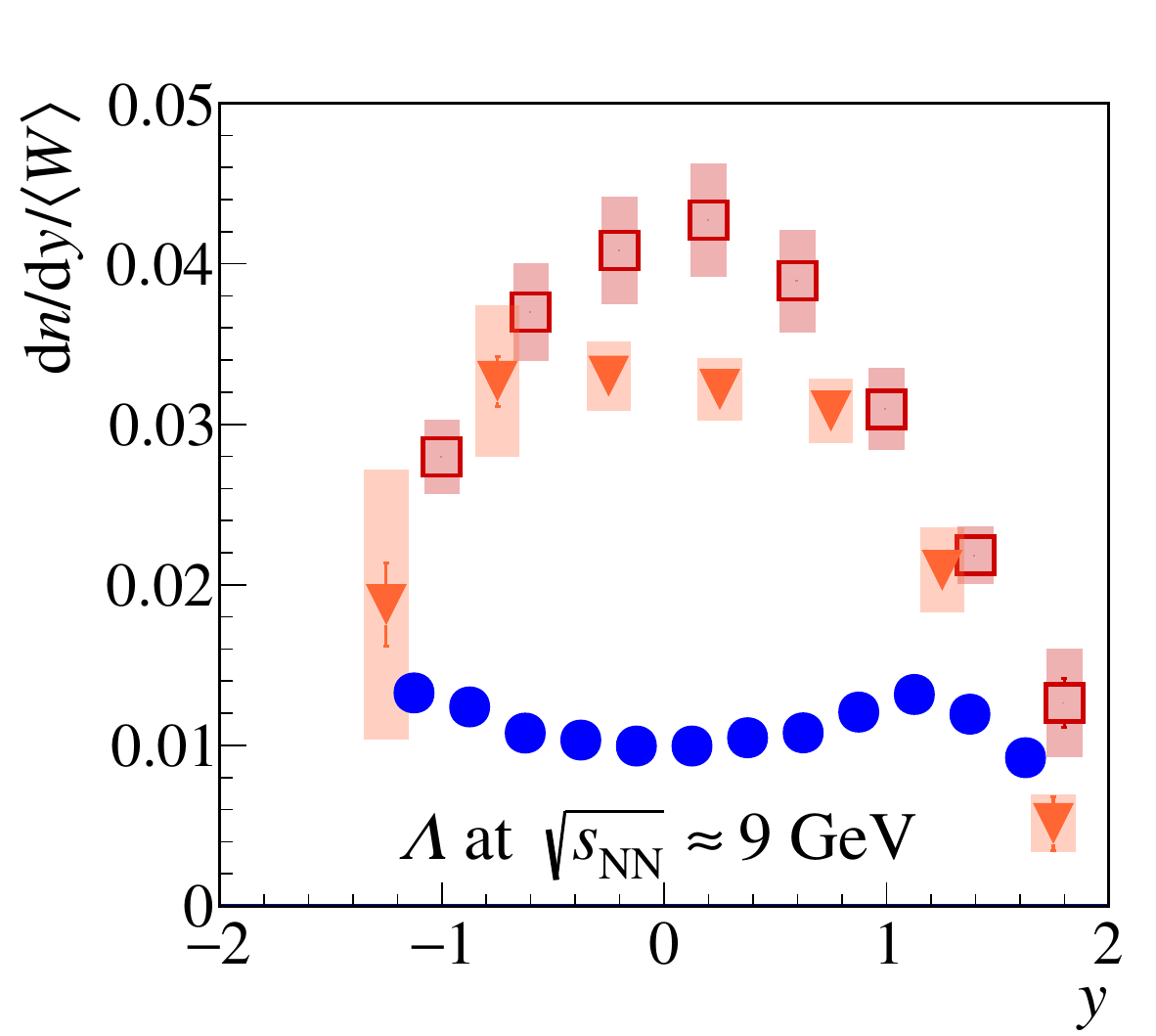} 
        \includegraphics[trim=0 0.0cm 0 1.5cm,clip,width = 0.301\linewidth]{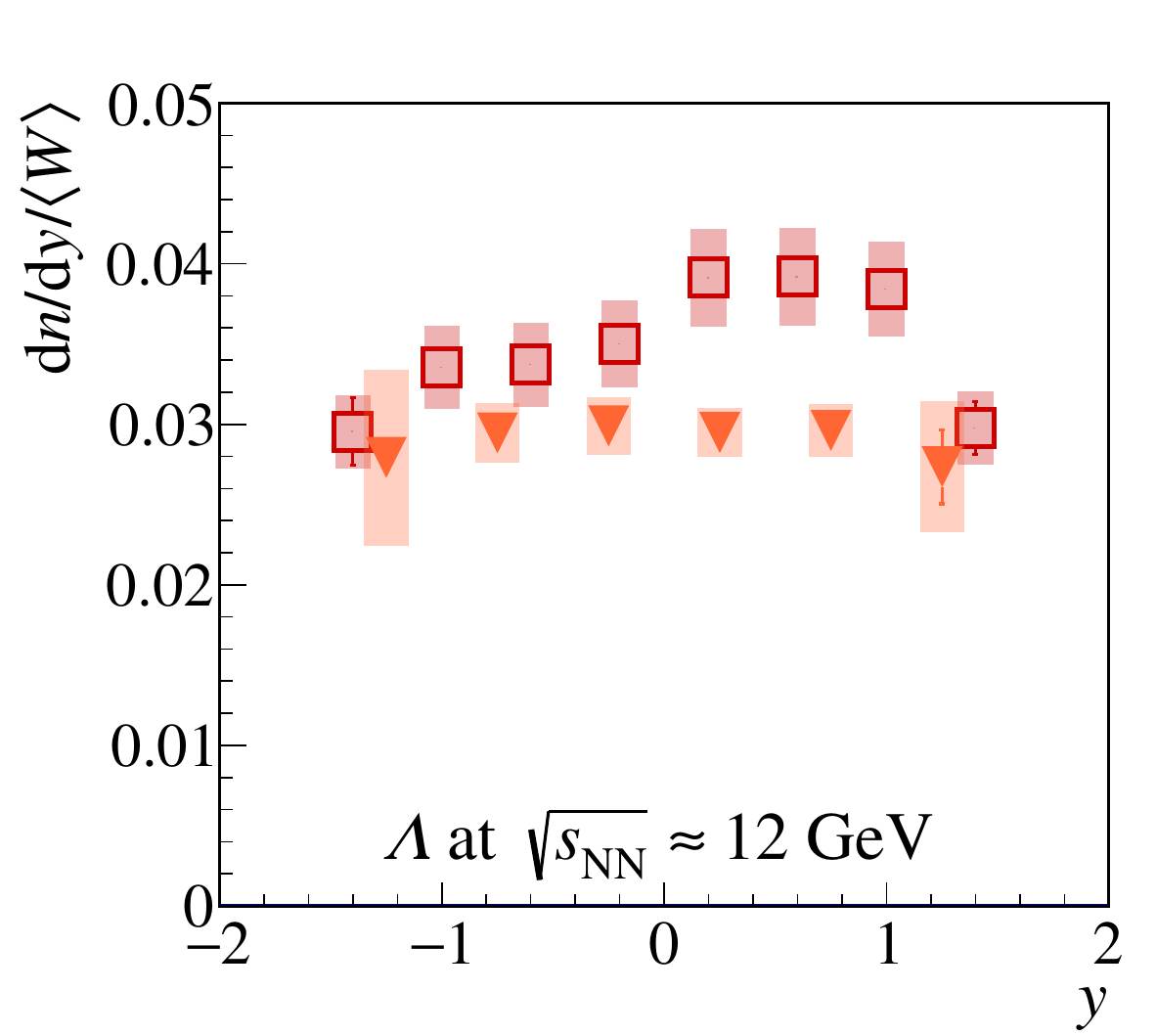} 
        \includegraphics[trim=0 0.0cm 0 1.5cm,clip,width = 0.301\linewidth]{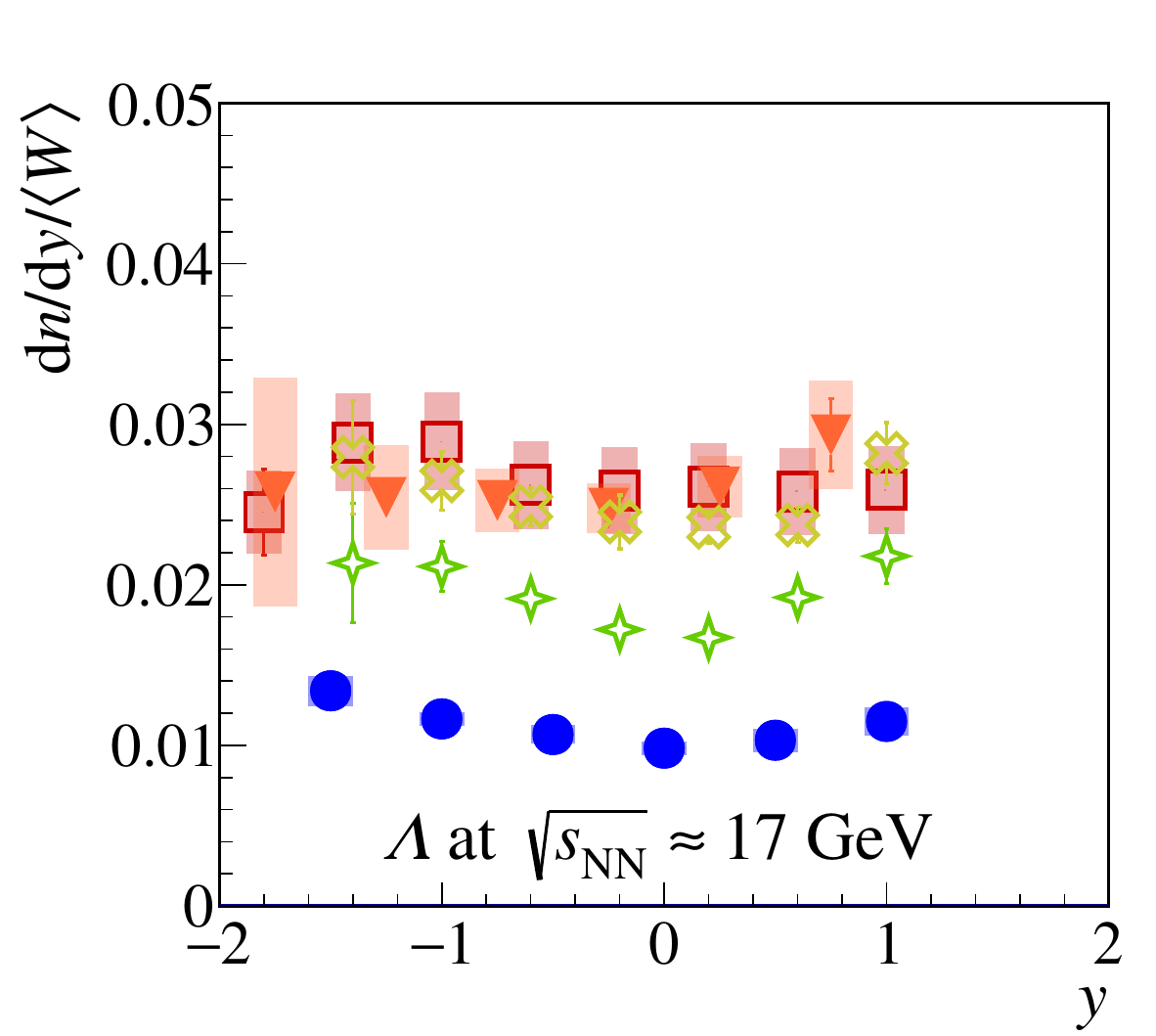} 
        \includegraphics[trim=0 0.0cm 0 1.5cm,clip,width = 0.301\linewidth]{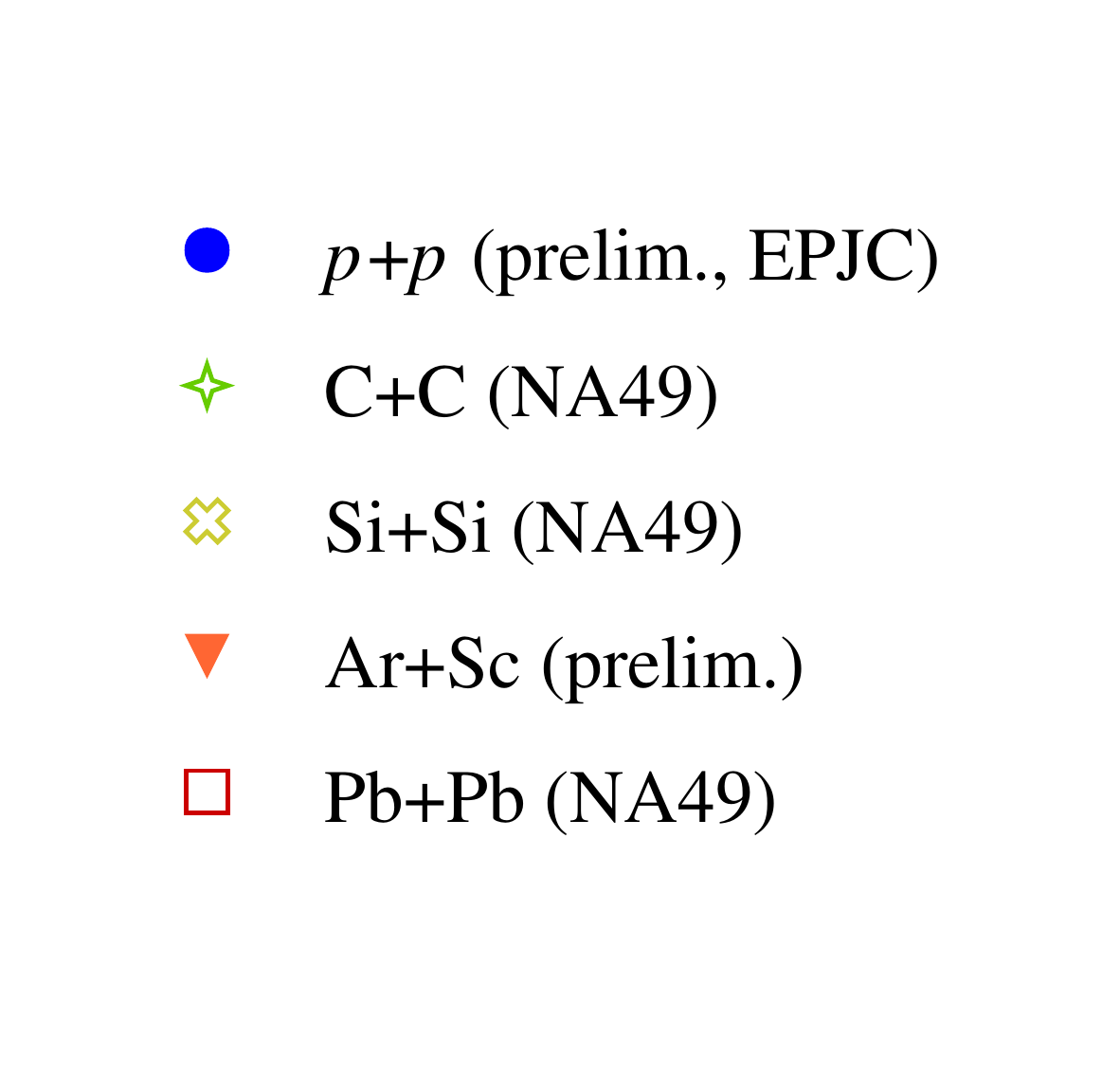} 
        \caption{Comparison of rapidity distributions $\mathrm{d}n/\mathrm{d}y$ of $\varLambda$ baryons produced in inelastic $p$+$p$ interactions and central C+C, Si+Si, Ar+Sc, and Pb+Pb collisions at $\sqrt{s_\mathrm{NN}} \approx 6, 8, 9, 12,$ and $17$~GeV. Statistical uncertainties are shown with error bars, and systematic uncertainties are shown as shaded bands. Published data from Refs.~\cite{paper:lambda_pp,paper:na49_pbpb_lambda,paper:na49_pbpb_lambda2}.}
        \label{figure:lambda}
        \end{center}
    \end{figure}

The rapidity distributions $\mathrm{d}n/\mathrm{d}y$ of $\pi^-$,  $K^+$, $p$, and $\varLambda$ produced in central collisions of different collision systems are presented in Figs.~\ref{figure:pi_minus}--\ref{figure:lambda}. All spectra are scaled by the mean number of wounded nucleons $\langle W \rangle$~\cite{paper:wounded}. Open markers in Figs.~\ref{figure:pi_minus}--\ref{figure:p} show values reflected symmetrically to $y=0$.

For $\pi^-$ (Fig.~\ref{figure:pi_minus}), a non-monotonic dependence of the amplitude of the spectra ($\mathrm{d}n/\mathrm{d}y/\langle W\rangle$ at $y\approx 0$) on the system size is observed: the amplitude reaches its maximum for the intermediate system -- Ar+Sc, and is lower for both lighter and heavier systems.

For $K^+$ (Fig.~\ref{figure:K}), the system-size dependence of the amplitude of the spectra is monotonic. Strangeness enhancement can be observed for intermediate and heavy systems (Ar+Sc, Xe+La, and Pb+Pb) compared to the light system (Be+Be). The spectra measured in Xe+La and Pb+Pb collisions agree within uncertainties, while the Ar+Sc yields are slightly lower (by up to 20\%). The Be+Be results are significantly below those for Ar+Sc, Xe+La, and Pb+Pb collisions.

The shape of the proton rapidity spectra (Fig.~\ref{figure:p}) is qualitatively different from that of mesons -- it displays a strong dependence on the collision energy and system size. It has been argued~\cite{paper:Panova} that a ``peak-dip-peak-dip'' transition as a function of energy exists only for the two-phase equation of state; thus, it can be viewed as one of the signatures of the onset of deconfinement. 
For Ar+Sc and Pb+Pb collisions, the single ``peak-dip'' transition is observed within the SPS energy range. 
However, the ``peak-dip-peak-dip'' irregularity is apparent for the Pb+Pb/Au+Au systems from a compilation of experimental data over the range of $\sqrt{s_\mathrm{NN}}$ from 2.4 to 17.3 GeV (Pb+Pb at SPS~\cite{na49_proton} and Au+Au at AGS~\cite{ags}). It remains to be elucidated whether similar behavior can also appear in Ar+Sc and Xe+La data over a broader energy range, e.g., including future data from the CBM experiment at FAIR~\cite{paper:CBM}. In the range of $\sqrt{s_\mathrm{NN}}$ from 6 to 17 GeV, no ``peak-dip'' transition is observed for Be+Be collisions.


$\varLambda$ hyperons, being both baryons and strange particles, are sensitive to two effects: baryon number transport and strangeness enhancement. As apparent in Fig.~\ref{figure:lambda}, spectra measured in Ar+Sc and Pb+Pb collisions approach each other with increasing collision energy.

\section{Energy and system size dependence of the strangeness-over-entropy production}
\label{sec:onset}

\begin{figure}[h]
         \includegraphics[width = 0.495\linewidth,trim={1.2cm 0cm 1.2cm 0cm}, clip]{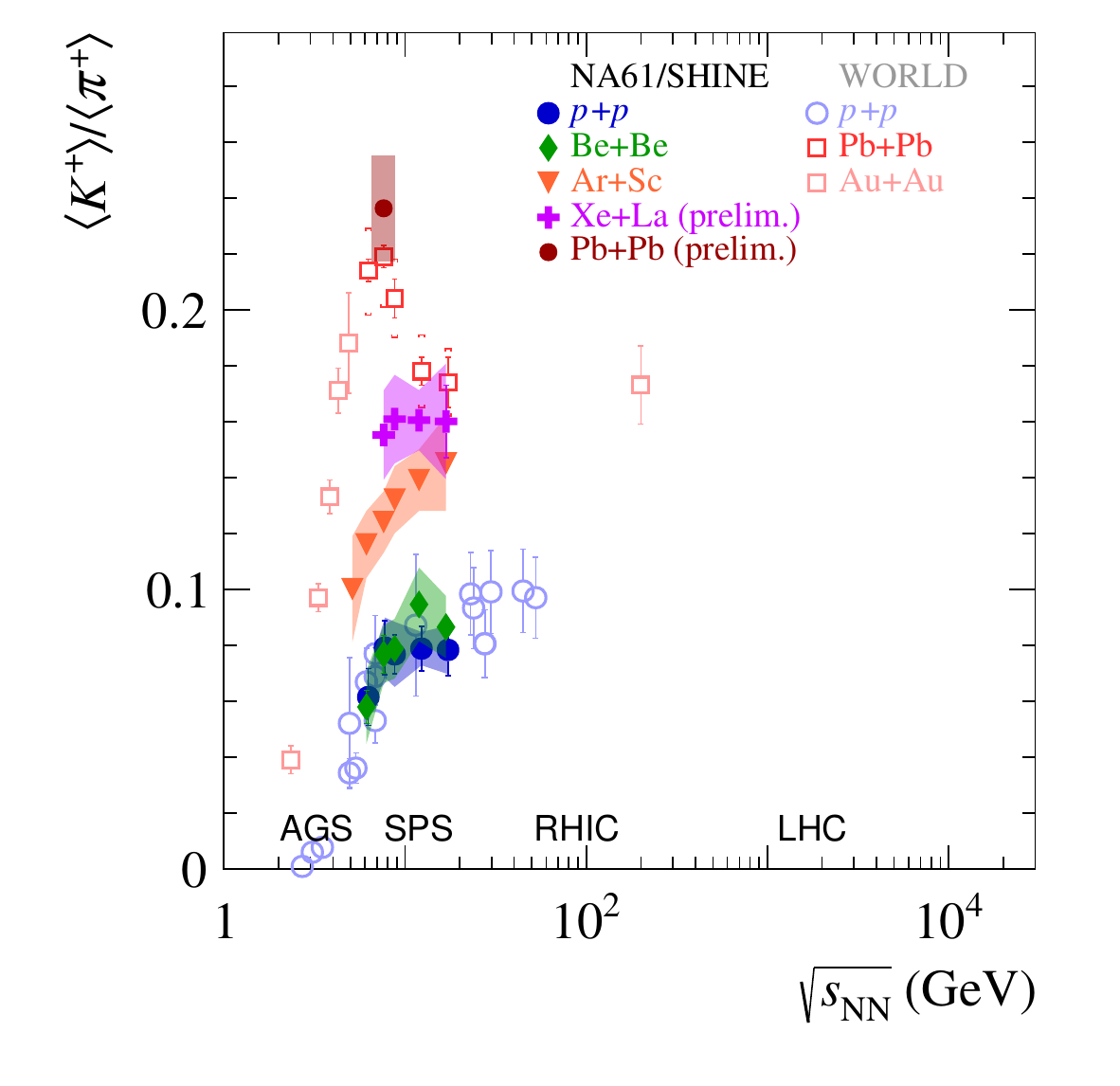}
          \includegraphics[width = 0.495\linewidth,trim={0.9cm 0cm 1.5cm 0cm}, clip]{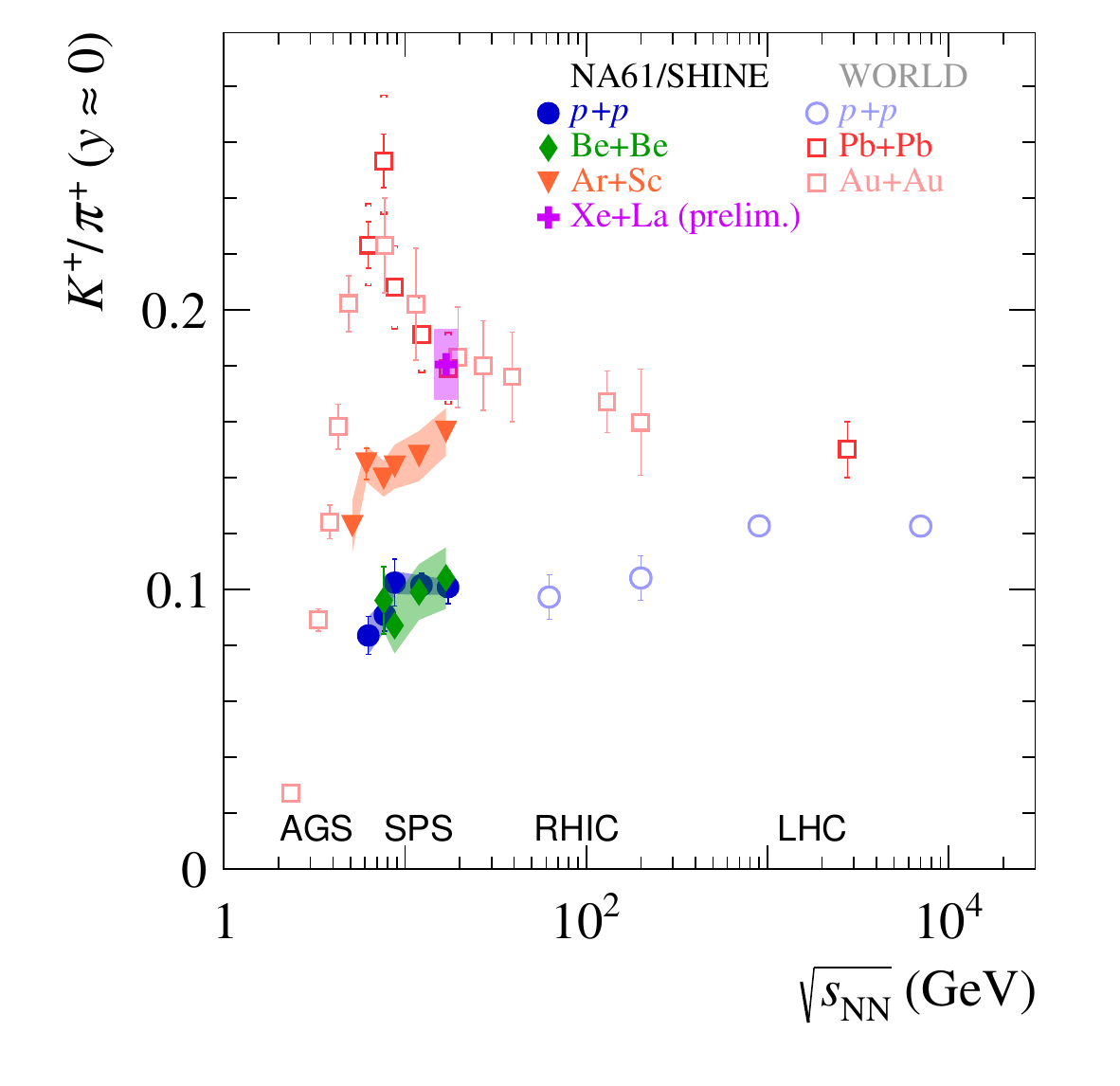}
          \caption{The energy dependence of the $\langle K^+\rangle/\langle \pi^+\rangle$ (\textit{left}) and $K^+/\pi^+~(y\approx0)$ (\textit{right}) for central Be+Be, Ar+Sc, Xe+La, Pb+Pb, and Au+Au collisions as
well as inelastic $p$+$p$ interactions. For Xe+La, $y = 0.4 - 0.6$ was used as mid-rapidity. For NA61/SHINE points, statistical uncertainties are shown as error bars and systematic as shaded bands. See Ref.~\cite{paper:na61_arsc_dedx} for references to published data. }
        \label{figure:horn}
    \end{figure}

According to the predictions of the SMES model~\cite{paper:smes}, the non-monotonic behavior of the strangeness-to-entropy ratio as a function of collision energy (``horn'') is one of the signatures of the onset of deconfinement. Experimentally, $\langle K^+\rangle/\langle \pi^+\rangle$ and $K^+/\pi^+~(y\approx0)$ are proportional to the strangeness-to-entropy ratio. From Fig.~\ref{figure:horn}, one can see that the ``horn'' is observed in central heavy-ion collisions (Au+Au~\cite{paper:AuAu1,paper:AuAu2, paper:AuAu3,paper:AuAu4,paper:AuAu5}, Pb+Pb~\cite{paper:PbPb1,paper:PbPb2,paper:PbPb3,paper:PbPb4}). However, no ``horn'' structure is apparent for Ar+Sc and Xe+La collisions. Notably, at the top energy the results for Ar+Sc and Xe+La collisions approach those for Pb+Pb, overlapping within uncertainties.

\begin{figure}[h]
    \centering
         \includegraphics[width = 0.500\linewidth]{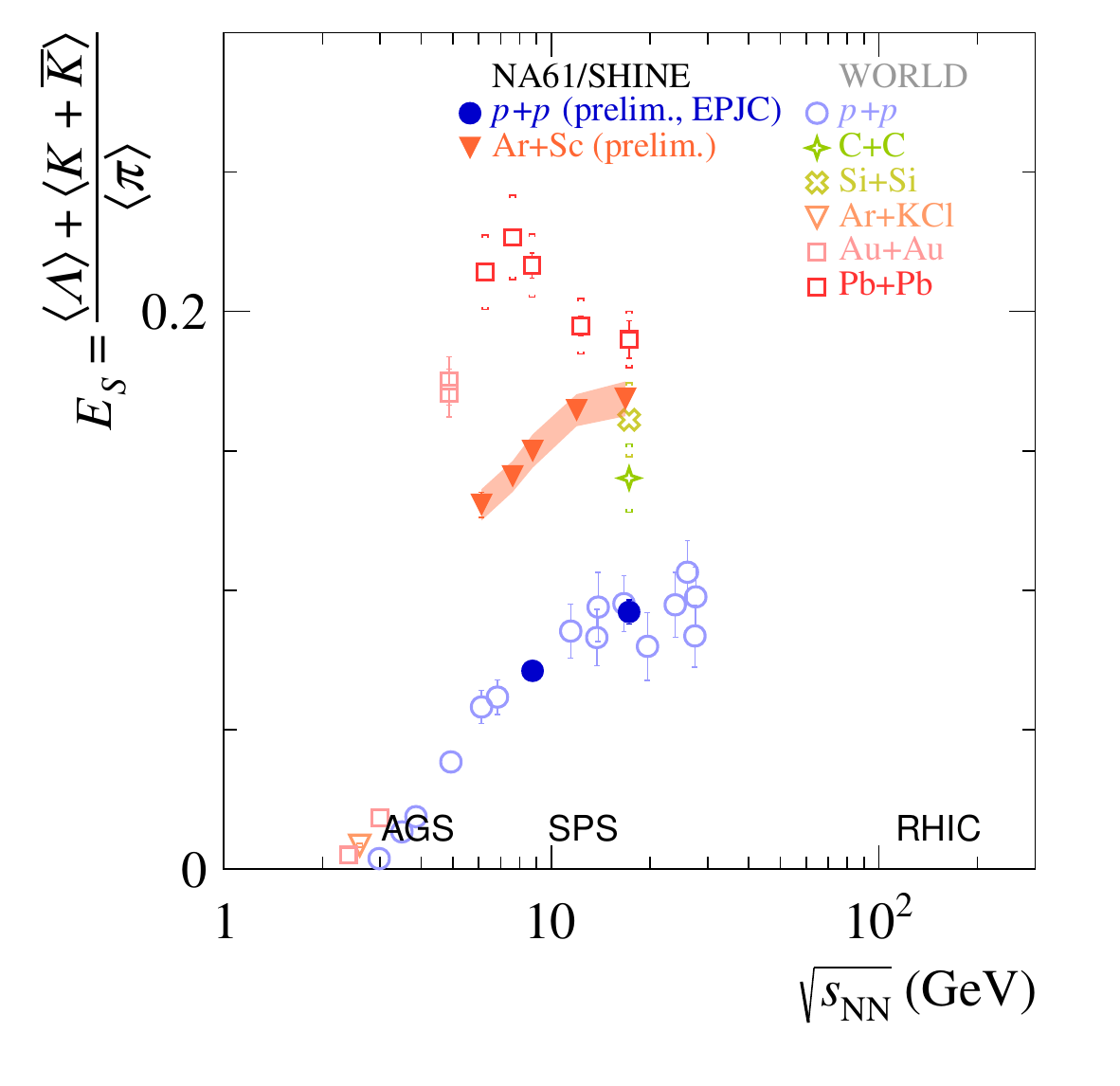}
          \caption{The energy dependence of the $E_S$ for central C+C, Si+Si, Ar+KCl, Ar+Sc, Pb+Pb, and Au+Au collisions as well as inelastic $p$+$p$ interactions. For NA61/SHINE points, statistical uncertainties are shown as error bars and systematic as shaded bands. See Ref.~\cite{talk:yuliia} for references to published data. }
        \label{figure:horn_lambda}
    \end{figure}

Figure~\ref{figure:horn_lambda} displays the $\sqrt{s_\mathrm{NN}}$ dependence of another measure proportional to the strangeness-to-entropy ratio -- $E_S = \frac{\langle \varLambda\rangle + \langle K+\overline K \rangle}{\langle\pi \rangle}$. The resulting picture is similar to what was observed for $\langle K^+\rangle/\langle \pi^+\rangle$ and $K^+/\pi^+~(y\approx0)$ ratios -- $E_S$ exhibits a non-monotonic horn structure for heavy systems, but monotonic dependence on the $\sqrt{s_\mathrm{NN}}$ for Ar+Sc collisions and lighter systems. However, at the top energy, results for Ar+Sc collisions approach those for Pb+Pb.

\begin{figure}
    \centering
    \includegraphics[width=0.500\linewidth]{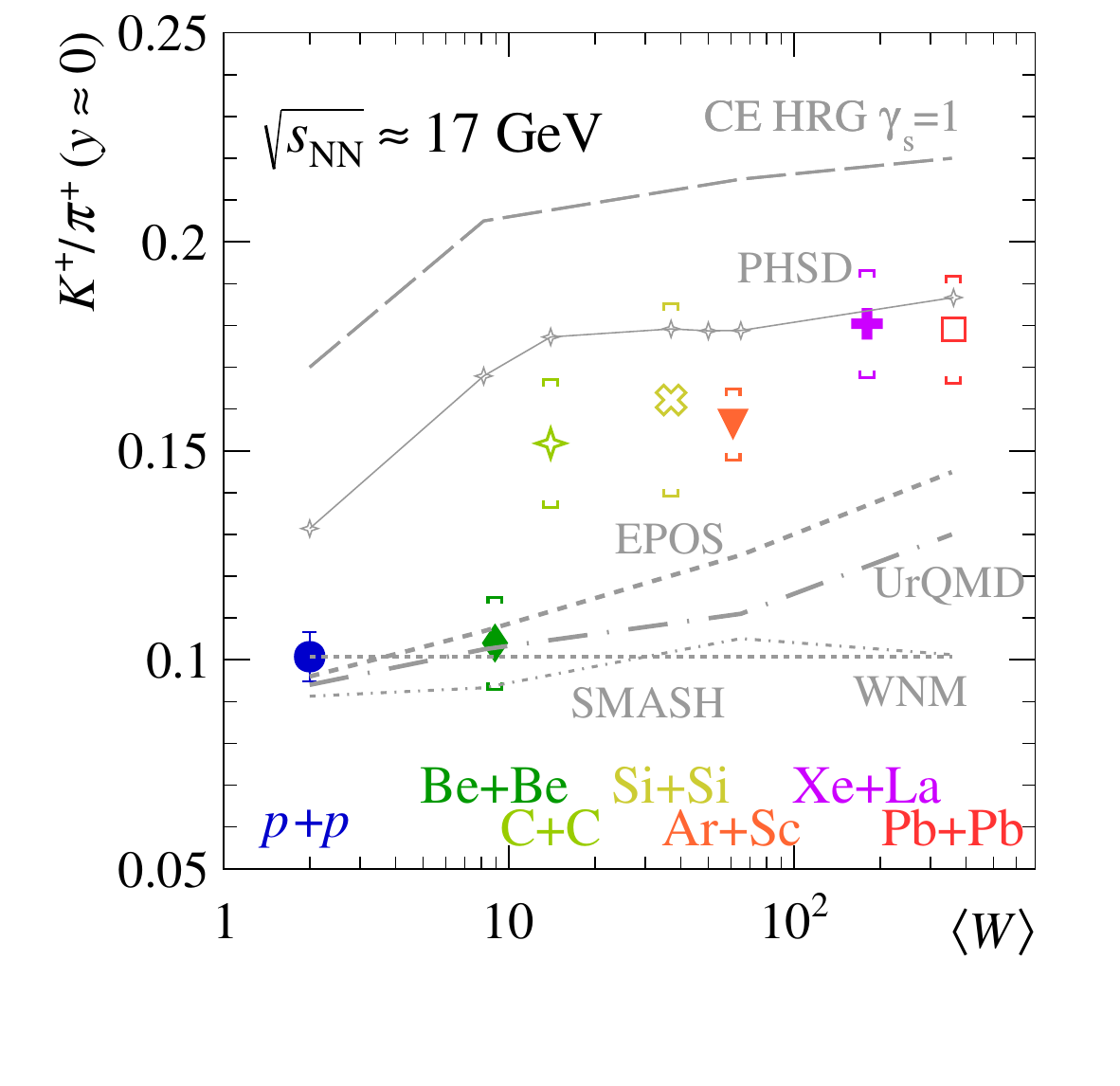}
    \caption{The system size dependence of $ K^+ / \pi^+  $ ratios at $y\approx 0$  measured  at $\sqrt{s_\mathrm{NN}} \approx 17$~GeV, showing values for inelastic $p$+$p$ interactions and central Be+Be, C+C, Si+Si, Ar+Sc, Xe+La, and Pb+Pb collisions. The system size is represented by the mean number of wounded nucleons. Statistical uncertainties are shown with bars and systematic uncertainties with square braces. Grey lines represent model simulations. The NA61/SHINE Xe+La result is preliminary and was obtained in the rapidity range $0.4 - 0.6$. See Ref.~\cite{paper:na61_arsc_dedx} for references to published data and model predictions.
}
        \label{figure:models}
\end{figure}

An exemplary comparison of the  system size dependence of measured $ K^+ / \pi^+  $ ratios at $y\approx 0$  with predictions of several theoretical models at $\sqrt{s_\mathrm{NN}} \approx 17$~GeV is shown in~Fig.~\ref{figure:models}. Experimental data on inelastic $p$+$p$ interactions
and central Be+Be, C+C, Si+Si, Ar+Sc, Xe+La, and Pb+Pb collisions are shown. None of the considered models describe the data. A further account of the failure of selected theoretical scenarios to describe the experimental strangeness-to-pion data in the SPS sector can be found in Ref.~\cite{maja}.\\
\newline

\clearpage

\bibliographystyle{ieeetr_n}
\bibliography{references}

\end{document}